\documentclass[preprint,superbib,floats]{revtex4}
\usepackage{graphicx,color}
\usepackage[centertags]{amsmath}
\usepackage{paralist}
\usepackage{tabularx}
\usepackage{delarray}
\usepackage{bm}
\usepackage[normalem]{ulem}
\usepackage{natbib}
\usepackage[english]{babel}

\begin{document}

\title{Cross-section geometry effects in the subband structure and spin-related properties of a HgTe/CdTe nanowire}

\date{\today}

\author{J. A. Budagosky}
\affiliation{Donostia International Physics Center (DIPC), Manuel de Lardizabal Pasealekua, E-20018 Donostia, Basque Country, Spain}

\begin{abstract}
By means of a multiband effective mass Hamiltonian, a theoretical characterization of the effect of the geometrical features of the confinement profile --in particular, a longitudinal groove-- on the subbands dispersion and spin-related properties of a rectangular HgTe/CdTe nanowire is presented. Through an external electric field applied perpendicular to the wire, the interplay of the induced Rashba spin splitting and these geometrical features is investigated. It is found that by exploiting this interplay a rich complexity of the subband structure arises, permitting  the generation and modulation of spin-polarized currents without magnetic fields. 
\end{abstract}

\maketitle

\section{Introduction}\label{introduction}

Research on the spin-related phenomenology in magnetic and semiconductor materials has grown considerably in the last years, primarily motivated by the increasing interest in the field of spintronics~\cite{Heinrich2000,Wolf2000,DasSarma2000,Papp2001,Zutic2004}. In order to understand properly the operation of these devices a detailed knowledge of the band structure and related aspects of the spin is of paramount  importance. 

One of the main advantages in the use of semiconductors in spintronics is the spin-orbit coupling (SOC) and the resulting spin splitting associated with these materials. In low-dimensional semiconductor systems, one can tune the spin splitting~\cite{Datta1990,Nitta1997,Koga2002} taking advantage of the presence of the Dresselhaus~\cite{Dresselhaus1955} and Rashba~\cite{Rashba1984} SOC. The Dresselhaus SOC arises in materials whose crystal structure lacks inversion symmetry (bulk inversion asymmetry, BIA) and leads to a spin splitting which depends on the electron wave vector, while Rashba SOC arises due to the absence of inversion symmetry of the confinement profile of a heterostructure~\cite{Rashba1984} (structural inversion asymmetry, SIA). The spin splitting is very important since it allows, for example, the control of the spin polarization by an electric field and the determination of the spin relaxation rate~\cite{Zutic2004}. The interference between the spin splitting due to Dresselhaus and Rashba SOC can lead to macroscopic effects, important for their potential applications~\cite{Fiederling1999,Pikus1995,Averkiev1999,Averkiev2002}. Additional complexities in the energy subband structure can be obtained in the presence of an external magnetic field. Furthermore, it is well known that the simultaneous presence of the SOC and external fields gives rise to a spatial variation in the distribution of the spin density --also known as spin texture-- for each subband. The \textit{spin current} concept is closely related to the spin texture because, experimentally, this current can be measured in terms of local variation of spin density~\cite{Khomitsky09}. The coherent transmission of information within electronic devices is the main goal considered when production, detection and manipulation of spin currents are investigated~\cite{Bratkovsky2008,Nichele2015}. 

Most of the investigations carried out in connection with SOC effects focus on 2D heterostructures and quantum wires (quasi 1D structures) with relatively simple shapes. However, the study of the effects of the SOC in more complicated structures, such as quantum wires with non-trivial cross-sections, is interesting by virtue of the new effects that may arise from these configurations. In this work some numerical results, obtained by means of an eight-band $\mathbf{k \cdot p}$ Hamiltonian, are shown in order to study in detail the effects of the SOC on the energy spectrum and spin polarization of a HgTe rectangular thin quantum wire embedded within a CdTe matrix. In addition, a longitudinal groove is included as part of the cross-section. As we will see later, this type of constriction has profound consequences on the subband structure and spin texture when an external electric field perpendicular to the nanowire is included.

We have chosen the HgTe/CdTe heterostructure as our model system for two reasons: first, HgTe is a negative gap material, i.e. the $\Gamma_6$ band, together with the $\Gamma_7$ band and the $\pm 3/2$-spin branches of the $\Gamma_8$ band, behave as valence bands, while the $\pm 1/2$-spin branches of the $\Gamma_8$ band behave as the conduction band. This feature makes HgTe a semimetal. This band inversion between $\Gamma_6$ and $\Gamma_8$ has profound consequences when the HgTe forms part of a heterostructure together with an insulator or a positive gap semiconductor, e.g. CdTe. The transition between an inverted (negative gap) subband structure and regular (positive gap) subband structure of a quantum well formed by the system CdTe/HgTe/CdTe, or a combination of their alloys, depends on the thickness of the well (e.g. for a HgTe/CdTe quantum well this critical thickness is about $~6.5$ nm). The most striking consequence of the particular alignment of the subband structure in HgTe quantum wells is the appearance of interfacial --or edge-- states~\cite{Johnson1988,Sengupta2013,Maciejko2011}. Second, together with its topological insulator (TI) character and tunability, the HgTe is a material with a quite large SOC, which makes it very interesting for spintronics. Furthermore, the lattice parameters of HgTe and CdTe are very similar, so the strain generated by the lattice mismatch between these materials is negligible. The latter will allow us to focus on the purely geometrical characteristics of the nanostructures studied in this paper. 

The paper is organized as follows: Section \ref{modeldescription} describes the theoretical background and the numerical method used in our calculations, and in section \ref{resultsanddiscussion} we show and discuss the obtained results.

\section{Description of the model}\label{modeldescription}
In this section, we will describe the implementation of the envelope function theory, together with the Hamiltonian used for studying the rectangular wire with and without the longitudinal groove.

\begin{figure}[t]
\includegraphics[trim=0cm 0cm 0cm 0cm, clip=true,width=0.4\textwidth]{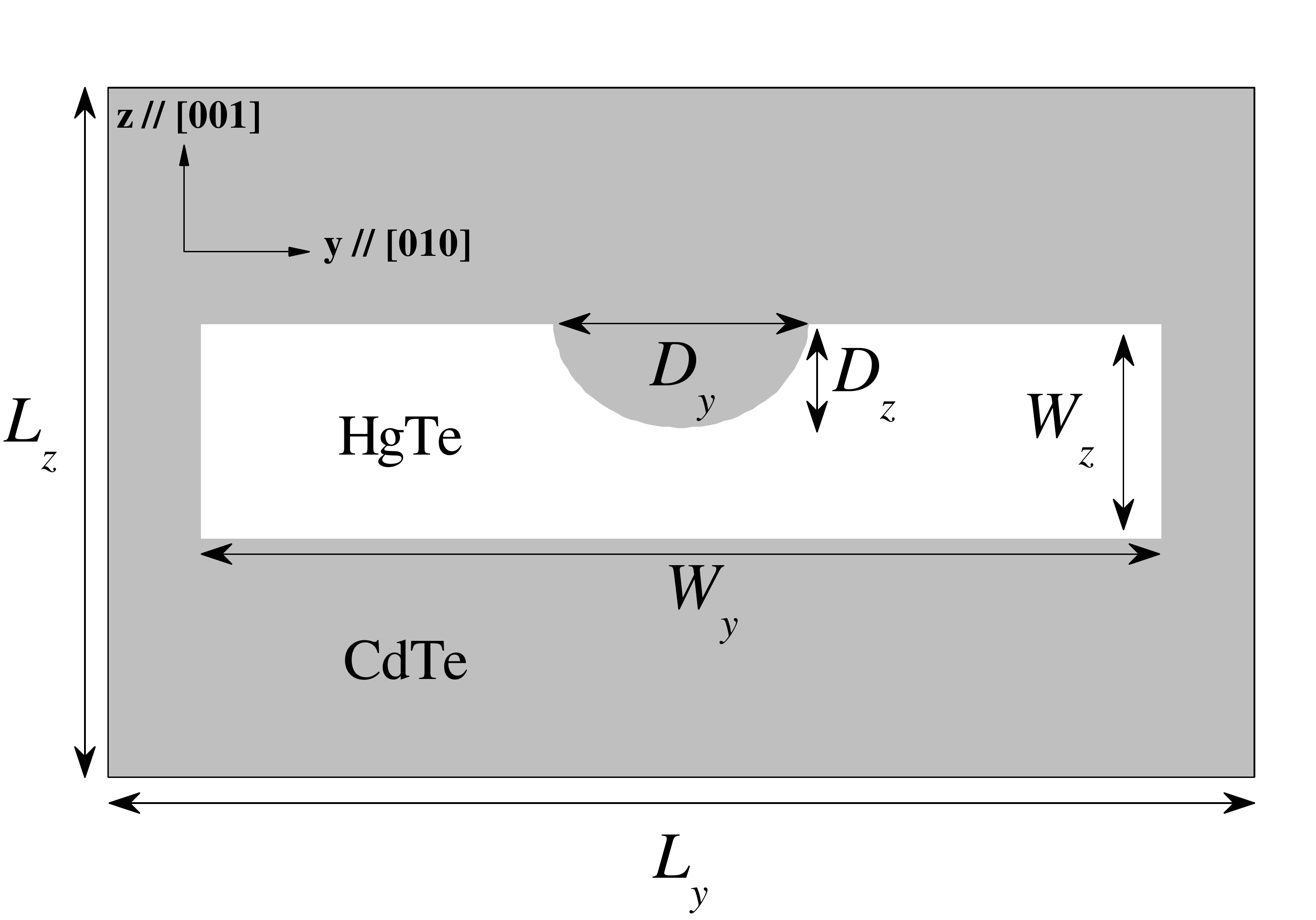}\\
\caption{Schematic model of one of the two HgTe/CdTe nanowires considered in our study. The other is the same that the shown here (rectangular shape) but without the groove.}\label{fig-1}
\end{figure}

The figure \ref{fig-1} shows the system under study: a HgTe rectangular thin quantum wire surrounded by a CdTe barrier. A longitudinal --semi-elliptical-- groove may be included in the upper facet of the wire cross-section. These types of shapes could be realizable, e.g., by etching or lithographic techniques~\cite{Snow1994,Cao2011}. The dimensions of the computational domain are defined by $L_{y}$ and $L_{z}$. A plane-wave expansion method is used in order to obtain the electronic structure of our system, which implies the use of periodic boundary conditions. In addition, an external uniform electric field, $\bm{\varepsilon}=(0,0,\varepsilon_z)$, is applied parallel to the $z$ axis.

We work in the framework of the Burt-Foreman (BF) envelope function theory~\cite{Foreman93,Foreman97,Burt1999}, employing an 8-band effective mass Hamiltonian that include BIA~\cite{Winkler2003,YanVoon2009}. The electric field is added to the diagonal elements of the Hamiltonian matrix. Since in this Hamiltonian the interaction between the conduction and valence bands is described explicitly, the Rashba-type spin splitting is automatically generated by including the asymmetry in the confining potential making the electric field different from zero. The basis of Bloch wave functions used to construct this Hamiltonian is:
\begin{eqnarray}\label{eq:basisbloch}
|u_1\rangle&=&|\frac{1}{2},+\frac{1}{2}\rangle=|S\uparrow\rangle \nonumber\\
|u_2\rangle&=&|\frac{1}{2},-\frac{1}{2}\rangle=|S\downarrow\rangle \nonumber\\
|u_3\rangle&=&|\frac{3}{2},+\frac{3}{2}\rangle=-\frac{1}{\sqrt{2}}|(X+iY)\uparrow\rangle \nonumber\\
|u_4\rangle&=&|\frac{3}{2},+\frac{1}{2}\rangle=-\frac{1}{\sqrt{6}}|(X+iY)\downarrow \rangle+\sqrt{\frac{2}{3}}|Z\uparrow \rangle \nonumber\\
|u_5\rangle&=&|\frac{3}{2},-\frac{1}{2}\rangle=\frac{1}{\sqrt{6}}|(X-iY)\uparrow \rangle+\sqrt{\frac{2}{3}}|Z\downarrow \rangle \nonumber\\
|u_6\rangle&=&|\frac{3}{2},-\frac{3}{2}\rangle=\frac{1}{\sqrt{2}}|(X-iY)\downarrow\rangle \nonumber\\
|u_7\rangle&=&|\frac{1}{2},+\frac{1}{2}\rangle=-\frac{1}{\sqrt{3}}|(X+iY)\downarrow \rangle-\frac{1}{\sqrt{3}}|Z\uparrow \rangle \nonumber\\
|u_8\rangle&=&|\frac{1}{2},-\frac{1}{2}\rangle=-\frac{1}{\sqrt{3}}|(X-iY)\uparrow \rangle+\frac{1}{\sqrt{3}}|Z\downarrow \rangle \nonumber\\
\end{eqnarray}
\noindent where $|X\rangle$, $|Y\rangle$, $|Z\rangle$ and $|S\rangle$ are the orbital wave functions of the top of the valence bands ($p_{x,y,z}$-type orbitals) and the bottom of the conduction band ($s$-type orbital), respectively. The symbols $\uparrow$ and $\downarrow$ denote spin-up and spin-down components. For the sake of clarity, we have divided the Hamiltonian into two parts: the zero-field Hamiltonian without BIA, $\hat{H}_{k}$, and a $8\times8$ matrix with the BIA-related parameters, $\hat{H}_{\text{BIA}}$:
\begin{widetext}
\begin{subequations}
\begin{eqnarray}\label{eq:hamiltonian8x8}
\hat{H}_{k}=
  \left( {\begin{array}{cccccccc}
   T_k & 0 & -\frac{P_0}{\sqrt{2}}k_+ & \sqrt{\frac{2}{3}}P_0k_z & \frac{P_0}{\sqrt{6}}k_- & 0 & -\frac{P_0}{\sqrt{3}}k_z & -\frac{P_0}{\sqrt{3}}k_-  \\
   0 & T_k & 0 & -\frac{P_0}{\sqrt{6}}k_+ & \sqrt{\frac{2}{3}}P_0k_z & \frac{P_0}{\sqrt{2}}k_- & -\frac{P_0}{\sqrt{3}}k_+ & \frac{P_0}{\sqrt{3}}k_z  \\
   -\frac{P_0}{\sqrt{2}}k_- & 0 &P_k+Q_k & S_k & R_k & 0 & -\frac{1}{\sqrt{2}}S_k & -\sqrt{2}R_k  \\
   \sqrt{\frac{2}{3}}P_0k_z & -\frac{P_0}{\sqrt{6}}k_- & S^{\dagger}_k &P_k-Q_k & M_k & R_k & \sqrt{2}Q_k & \sqrt{\frac{3}{2}}\Sigma_k  \\
   \frac{P_0}{\sqrt{6}}k_+ & \sqrt{\frac{2}{3}}P_0k_z & R^{\dagger}_k & M^{\dagger}_k & P_k-Q_k & -S^T_k & \sqrt{\frac{3}{2}}\Sigma^*_k & -\sqrt{2}Q_k  \\
   0 & \frac{P_0}{\sqrt{2}}k_+ & 0 & R^{\dagger}_k & -S^*_k & P_k+Q_k & \sqrt{2}R^{\dagger}_k & -\frac{1}{\sqrt{2}}S^*_k  \\
   -\frac{P_0}{\sqrt{3}}k_z & -\frac{P_0}{\sqrt{3}}k_- & -\frac{1}{\sqrt{2}}S^{\dagger}_k & \sqrt{2}Q_k & \sqrt{\frac{3}{2}}\Sigma^T_k & \sqrt{2}R_k & P_k-\Delta_{SO} & M_k  \\
   -\frac{P_0}{\sqrt{3}}k_+ & \frac{P_0}{\sqrt{3}}k_z & -\sqrt{2}R^{\dagger}_k & \sqrt{\frac{3}{2}}\Sigma^{\dagger}_k & -\sqrt{2}Q_k & -\frac{1}{\sqrt{2}}S^T_k & M^{\dagger}_k & P_k-\Delta_{SO}  \\
  \end{array} } \right)\text{,}\nonumber\\
  \end{eqnarray}
\begin{eqnarray}
\hat{H}_{\text{BIA}}=
  \left( {\begin{array}{cccccccc}
  0 & 0 & \frac{1}{2\sqrt{2}}A_k & \frac{1}{\sqrt{6}}O_k & \frac{1}{2\sqrt{6}}A^*_k & \frac{1}{3\sqrt{2}}N_k & -\frac{i}{2\sqrt{3}}D_k & -\frac{1}{2\sqrt{3}}L_k \\
  0 & 0 & -\frac{1}{3\sqrt{2}}N_k & \frac{1}{2\sqrt{6}}A_k & -\frac{1}{\sqrt{6}}O_k & \frac{1}{2\sqrt{2}}A^*_k & \frac{1}{2\sqrt{3}}L^*_k & \frac{i}{2\sqrt{3}}D_k \\
  \frac{1}{2\sqrt{2}}A^{\dagger}_k & \frac{1}{3\sqrt{2}}N^{\dagger}_k & 0 & -\frac{1}{2}C_0k_+ & C_0k_z & -\frac{\sqrt{3}}{2}C_0k_- & \frac{1}{2\sqrt{2}}C_0k_+ & \frac{1}{\sqrt{2}}C_0k_z \\
  \frac{1}{\sqrt{6}}O^{\dagger}_k & \frac{1}{2\sqrt{6}}A^{\dagger}_k & -\frac{1}{2}C_0k_- & 0 & \frac{\sqrt{3}}{2}C_0k_+ & -C_0k_z & 0 & -\frac{\sqrt{3}}{2\sqrt{2}}C_0k_+ \\
  \frac{1}{2\sqrt{6}}A^T_k & -\frac{1}{\sqrt{6}}O^{\dagger}_k & C_0k_z & \frac{\sqrt{3}}{2}C_0k_- & 0 & -\frac{1}{2}C_0k_+ & -\frac{\sqrt{3}}{2\sqrt{2}}C_0k_- & 0 \\
  \frac{1}{3\sqrt{2}}N^{\dagger}_k & \frac{1}{2\sqrt{2}}A^T_k & -\frac{\sqrt{3}}{2}C_0k_+ & -C_0k_z & -\frac{1}{2}C_0k_- & 0 & \frac{1}{\sqrt{2}}C_0k_z & -\frac{1}{2\sqrt{2}}C_0k_- \\
  \frac{i}{2\sqrt{3}}D^{\dagger}_k & \frac{1}{2\sqrt{3}}L^T_k & \frac{1}{2\sqrt{2}}C_0k_- & 0 & -\frac{\sqrt{3}}{2\sqrt{2}}C_0k_+ & \frac{1}{\sqrt{2}}C_0k_z & 0 & 0 \\
  -\frac{1}{2\sqrt{3}}L^{\dagger}_k & \frac{i}{2\sqrt{3}}D^{\dagger}_k & \frac{1}{\sqrt{2}}C_0k_z  & -\frac{\sqrt{3}}{2\sqrt{2}}C_0k_- & 0 & -\frac{1}{2\sqrt{2}}C_0k_+ & 0 & 0 \\
  \end{array} } \right)\nonumber\\
  \label{eq:hamiltonian8x8BIA}
  \normalsize
\end{eqnarray}
\end{subequations}
\end{widetext}

\noindent where $k_{\pm}=k_x{\pm}i k_y$, $k_y=-i\partial/\partial y$ and $k_z=-i\partial/\partial z$. The superscript $^\dagger$ refers to hermitian conjugation, $^T$ means transpose and $^*$ means complex conjugation. In the $[001]$-oriented Hamiltonian, the terms $T_k$, $P_k$, $Q_k$, $S_k$, $\Sigma_k$, $R_k$ and $M_k$ in Eq.~\ref{eq:hamiltonian8x8} are given by:
\begin{subequations}
\begin{eqnarray}
&&T_k=E_c+\frac{\hbar^2}{2m_0}\left(k_x\gamma'k_x+k_y\gamma'k_y+k_z\gamma'k_z\right)\\
&&P_k=E_v-\frac{\hbar^2}{2m_0}\left(k_x\gamma'_1k_x+k_y\gamma'_1k_y+k_z\gamma'_1k_z\right)\\
&&Q_k=-\frac{\hbar^2}{2m_0}\left(k_x\gamma'_2k_x+k_y\gamma'_2k_y-2k_z\gamma'_2k_z\right)\\
&&R_k=\sqrt{3}\frac{\hbar^2}{2m_0}\left[ k_x\gamma'_2k_x - k_y\gamma'_2k_y \right. \nonumber\\
&&~~~~~~~~~~~~~~~~~~\left.-i\left(k_x\gamma'_3k_y+k_y\gamma'_3k_x\right)\right]\\
&&S_k=\sqrt{3}\frac{\hbar^2}{2m_0}\left( k_-\gamma'_3k_z + k_z\gamma'_3k_- \right) \nonumber\\
&&~~~~~~~~~~~~~~~~~~-\sqrt{3}\frac{\hbar^2}{2m_0}\left( k_z\chi k_- - k_-\chi k_z \right)\\
&&\Sigma_k=\sqrt{3}\frac{\hbar^2}{2m_0}\left( k_-\gamma'_3k_z + k_z\gamma'_3k_- \right) \nonumber\\
&&~~~~~~~~~~~~~~~~~~+\sqrt{\frac{1}{3}}\frac{\hbar^2}{2m_0}\left( k_z\chi k_- - k_-\chi k_z \right)\\
&&M_k=-2\frac{\hbar^2}{2m_0}\left( k_z\chi k_- - k_-\chi k_z \right)~\text{.}
\end{eqnarray}
\end{subequations}\label{eq:hamiltonian8x8elements}

Here, $E_c$, $E_v$ and $\Delta_{SO}$ are the conduction band edge, the valence band edge and the spin-orbit split-off parameters, respectively, while $\chi$ is an anisotropy parameter and $P_0$ is the Kane momentum matrix element. In addition, $\gamma'$ and $\gamma'_i$ are the conduction band parameter and the valence band parameters, respectively. The values of these have been chosen properly to describe the coupling to the remote bands. On the other hand, the terms $A_k$, $O_k$, $N_k$, $D_k$, $L_k$ are given by:
\begin{subequations}
\begin{eqnarray}
&&A_k=k_-B^+_{8v}k_z+k_zB^+_{8v}k_-\\
&&O_k=k_xB^-_{8v}k_x-k_yB^-_{8v}k_y \nonumber \\
&&~~~~+i\left(k_xB^+_{8v}k_y+k_yB^+_{8v}k_x\right)\\
&&N_k=k_xB^-_{8v}k_x+k_yB^-_{8v}k_y-2k_zB^-_{8v}k_z \\
&&D_k=k_xB_{7v}k_y+k_yB_{7v}k_x\\
&&L_k=k_+B_{7v}k_z+k_zB_{7v}k_+
\end{eqnarray}
\end{subequations}\label{eq:hamiltonian8x8BIAelements}

\noindent where $B_{8v}^{\pm}$ and $B_{7v}$ are the BIA band parameters related with the terms cuadratic in $k$ in Eq.~\eqref{eq:hamiltonian8x8BIA}. The terms linear in $k$ are weighted by the parameter $C_0$. All parameters described above were taken from Ref.~\onlinecite{Novik2005} and~\onlinecite{Winkler2012}. Except for $P_0$, $B^{\pm}_{8v}$ and $B_{7v}$, the rest of the parameters are position dependent. This dependence is described explicitly as:
\begin{equation}\label{eq:positiondependence}
f(\mathbf{r}) = f^{\text{CdTe}} + (f^{\text{HgTe}}-f^{\text{CdTe}})\alpha(\mathbf{r})~\text{,}
\end{equation}
\noindent being $f^{\text{HgTe}}$ ($f^{\text{CdTe}}$) the value of the parameter in the wire (barrier) and $\alpha(\mathbf{r})$ a characteristic function that defines the shape of the cross-section of the nanowire (is set as unity within the wire and zero in the barrier). This is calculated numerically in a real-space grid.

Note that we have made the operator form of the wave vector explicit for all components (even along the wire axis) so these expressions are valid also for three-dimensional quantization. In addition, all the elements of the matrices \eqref{eq:hamiltonian8x8} and \eqref{eq:hamiltonian8x8BIA} that depend linearly on the wave vectors are treated in a symmetrized fashion. Finally, the Hamiltonian can be expressed as:
\begin{equation}\label{eq:hamiltonianfull}
\hat{H}=\hat{H}_k+\hat{H}_{\text{BIA}}+\hat{I}_{8\times8}(e\varepsilon_z z)\text{,}
\end{equation}
\noindent where $\hat{I}_{8\times8}$ is the $8\times8$ unit matrix and $\varepsilon_z$ is the external electric field strength applied parallel to $z$.

\subsection{Plane-wave expansion method}

We have implemented the above Hamiltonian for our problem using a plane-wave expansion within the first Brillouin zone --as this is demanded by the exact envelope function theory~\cite{YanVoon2009}. The unit-cell dimensions have been chosen to maintain a balance between maximizing the efficiency of the calculation and minimizing the coupling with the rest of the periodic array, at least with regard to the conduction and valence subbands considered here. Note that the eigenvalue problem using the Hamiltonian of Eq. \eqref{eq:hamiltonianfull} should be written formally in Fourier representation and not in real space. The characteristic function $\alpha(\mathbf{r})$ in Eq.~\eqref{eq:positiondependence} is transformed accordingly using a fast Fourier transform (FFT) routine.  

To start, we write the electron wave function in the material as an eight-component spinor:
\begin{equation}\label{eq:8compspinor}
|\psi_{k_x}(\bm{r})\rangle = e^{ik_xx}\sum_{\alpha=1}^{8}\chi_{\alpha}(\bm{r})|u_{\alpha}\rangle~~~\text{,}
\end{equation}

\noindent where $\bm{r}=(y,z)$ and $\chi_{\alpha}(\bm{r})$ is an envelope function associated with the slowly varying Bloch function of the bulk material. The first step in the plane-wave expansion method consists in expanding the envelope function as a linear combination of plane waves,
\begin{equation}\label{eq:envelopepw}
\chi_{\alpha}(\bm{r}) = \frac{1}{\sqrt{\Omega}}\sum_{\bm{q}}A_{\alpha\bm{q}}
e^{i\bm{q}\cdot \bm{r}}~~~\text{,}
\end{equation}
\noindent being $\bm{q}=\left(2\pi m_y/L_y,2\pi m_z/L_z\right)$, $\Omega=L_yL_z$ the area of the unit-cell and $A_{\alpha\bm{q}}$ a set of complex coefficients to be determined. Here, $m_i=-(M_i-1)/2,...,+(M_i-1)/2$, being $M_i$ the total number of plane waves along direction $i$. For a given value of $k_x$ and using standard diagonalization techniques~\cite{lapack1999}, the energy levels and eigenfunctions of the nanowire are found by solving the matrix eigenvalue problem of dimension $N\times N$ (with $N=8\times M_{y}\times M_{z}$), obtained from the differential equation that results from introducing Eq.~\eqref{eq:8compspinor} into Eq.~\eqref{eq:hamiltonianfull}:
\begin{equation}\label{eq:eigenvalueproblem}
\sum_{\beta=1}^8\sum_{\bm{q}}h_{\alpha\beta}(\bm{q'}.
\bm{q},k_x)A_{\beta\bm{q}}=
EA_{\alpha\bm{q'}}~~~\text{,}
\end{equation}

In order to set up the matrix of Eq.~\eqref{eq:eigenvalueproblem} we must first evaluate the matrix elements $h_{\alpha\beta}(\bm{q'},\bm{q},k_x)$ linking plane-wave basis states of wave vectors $\bm{q'}$ and $\bm{q}$:
\begin{equation}\label{eq:eigenvaluematrixelement}
h_{\alpha\beta}(\bm{q'},\bm{q},k_x)=
\frac{1}{\Omega}\int_{\Omega}e^{-i\bm{q'}\cdot \bm{r}}\hat{H}_{\alpha\beta}e^{i\bm{q}\cdot \bm{r}}d^2\bm{r}~~~\text{.}
\end{equation} 

The evaluation of $h_{\alpha\beta}(\bm{q'},\bm{q},k_x)$  is greatly facilitated if we take into account that the elements $\hat{H}_{\alpha\beta}$ of the $8\times8$ Hamiltonian matrix are expressed as linear combinations of elements of the form:

\begin{subequations}
\begin{eqnarray}\label{eq:typeofmatrixelement}
\zeta_1&=&f(\bm{r})~\text{,}\\
\zeta_2&=&f(\bm{r})k_n=k_nf(\bm{r})~\text{and}\\
\zeta_3&=&k_mf(\bm{r})k_n
\end{eqnarray}
\end{subequations}

\noindent with $n,m=y,z$. The elements $\zeta_i$ are evaluated in the basis of plane waves in the same way as in Eq.~\eqref{eq:eigenvaluematrixelement}, but replacing $\hat{H}_{\alpha\beta}$ by $\zeta_i$:
\begin{equation}\label{eq:eigenvaluepartialmatrixelement}
\zeta_i(\bm{q'},\bm{q})=
\frac{1}{\Omega}\int_{\Omega}e^{-i\bm{q'}\cdot \bm{r}}\zeta_ie^{i\bm{q}\cdot \bm{r}}d^2\bm{r}~~~\text{.}
\end{equation}

By making the substitutions 
\begin{eqnarray}\label{eq:differentialoperators}
k_nf(\bm{r}) &\rightarrow& -\frac{i}{2}\left( \partial_nf(\bm{r}) + f(\bm{r})\partial_n \right)~\text{,}\nonumber\\
k_mf(\bm{r})k_n &\rightarrow& -\partial_mf(\bm{r})\partial_n~\text{,}\nonumber\\
\end{eqnarray}
\noindent and integrating in Eq.~\eqref{eq:eigenvaluepartialmatrixelement}, the resulting expressions for the elements $\zeta_i(\bm{q'},\bm{q})$ are:
\begin{subequations}
\begin{eqnarray}
&&\zeta_1(\bm{q'},\bm{q}) = f^{\text{CdTe}}\delta_{\bm{q'},\bm{q}} + \frac{\Delta f}{\Omega}\tilde{\alpha}(\bm{q'}-\bm{q}) \\
&&\zeta_2(\bm{q'},\bm{q}) = \frac{\left( q'_n + q_n \right)}{2}\zeta_1(\bm{q'},\bm{q})\text{,}\\
&&\zeta_3(\bm{q'},\bm{q}) = q'_mq_n\zeta_1(\bm{q'},\bm{q})\text{,}
\end{eqnarray}\label{eq:eigenvaluepartialmatrixelement1}
\end{subequations}
\noindent being $\Delta f = f^{\text{HgTe}} - f^{\text{CdTe}}$ and $\tilde{\alpha}$ the characteristic function $\alpha(\bm{r})$ in Fourier representation, evaluated at $\bm{q'}-\bm{q}$. 
Once the elements $\zeta_i(\bm{q'},\bm{q})$ are evaluated, obtaining the matrix elements $h_{\alpha\beta}(\bm{q'},\bm{q},k_x)$ is straightforward. 

\subsection{Spin polarization and transport}

For the study of spin and charge transport in our system, we calculate the charge and spin-polarized conductance --we restrict ourselves to the ballistic regime-- along the axis of the wire as a function of the Fermi level. We assume a sufficiently long wire connected at both ends with two reservoirs with chemical potentials $\mu_S$ (source) and $\mu_D$ (drain). The external bias that generates the current is assumed to be created by a difference between these chemical potentials, $eV_c=\mu_S-\mu_D$. Starting from the definition given in Refs.~[\onlinecite{Pershin2004}] and~[\onlinecite{Serra2005}], at very low temperature and in the limit of small bias ($\mu_S\approx\mu_D$), the ballistic charge conductance can be re-written in terms of the wave vector $k_x$ as,
\begin{equation}\label{eq:chargeconductance}
G(E_f)=\frac{e^2}{h}\sum_{n,s}
\int^{+\infty}_{-\infty}v_{n,s}\Theta[v_{n,s}]
\delta(E_{n,s}(k_x)-E_f)dk_x~\text{,}
\end{equation}
\noindent where $v_{n,s}$ is the electron group velocity, $v_{n,s}(k_x)=\partial{E_{n,s}(k_x)}/\left(\hbar\partial{k_x}\right)$, for propagation along the wire with energy $E_{n,s}(k_x)$, $\Theta$ is the Heaviside function (the sign of $v_{n,s}$ defines the sign of the bias and, for instance, the direction of current propagation) and $E_f$ is the Fermi level. Here, $n$ is
the orbital quantum number --running from the lowest energy
subband to the highest-- while $s=\pm 1$ labels the spin branch of  the $n$'th subband. Finally, for numerical purposes, the $\delta(E_{n,s}(k_x)-E_f)$ function is approximated by a narrow Gaussian.

On the other hand, by numerical diagonalization of the Hamiltonian, it is possible to take a look at the spin polarization described by a momentum-dependent vector field, $\bm{S}(k_x)=\left(\langle S_x\rangle,\langle S_y\rangle,
\langle S_z\rangle\right)$, where $S_m$ are the Cartesian components of the spin matrix operator in the eight-band representation~\cite{Abolfath2001,Dargys2007}.

In that sense, from the obtained spectrum, we can calculate the average spin components of each $s=\pm 1$ spin branch per $n$'th subband along the $i=x,y,z$ directions, 
\begin{equation}\label{eq:spinexpectationvalues}
\langle S_i\rangle_{n,s}=\int_{\Omega}d^2\bm{r}
\psi^*_{n,s,k_x}(\bm{r})S_{i}\psi_{n,s,k_x}(\bm{r})
~~\text{.}
\end{equation}

Thus, following~[\onlinecite{Tokatly2008,Sun2008,Grigoryan2009}], for a given Fermi level $E_f$ the spin-polarized current related to components $\langle S_i\rangle_{n,s}$ can be written as
\begin{eqnarray}\label{eq:spinconductance}
J^{i}_{x}(E_f)&=&\frac{e^2}{h}\sum_{n,s}\int^{+\infty}_{-\infty}
\langle S_i\rangle_{n,s}v_{n,s}\times \nonumber\\
&&\Theta[v_{n,s}]\delta(E_{n,s}(k_x)-E_f)dk_x
~\text{,}\nonumber\\
\end{eqnarray}
\noindent which is basically the same expression as for the charge conductance [Eq.~\eqref{eq:chargeconductance}], but including the average spin polarization in the integrand.  Finally, it is important to recognize that, due to the SOC, the label $s$ is no longer a good quantum number. Nevertheless, we kept using it in some cases along this work for the sole purpose of differentiating between the lower ($s = -1$) and higher ($s = +1$) spin-branches of the same $n$'th subband.

\section{Numerical results and discussion}\label{resultsanddiscussion}

For the numerical calculations, we have used the following parameters: $L_y=100$ nm, $L_z=20$ nm, $W_y=80$ nm, $D_y=20$ nm and $D_z=2.5$ nm. The center of coordinates of our system ($y=0,z=0$) is located at the center of the nanowire.  Through the paper, we considered several values for the thickness, $W_z$. 

\begin{figure}[t]
\includegraphics[trim=0cm 0cm 0cm 0cm, clip=true,width=0.5\textwidth]{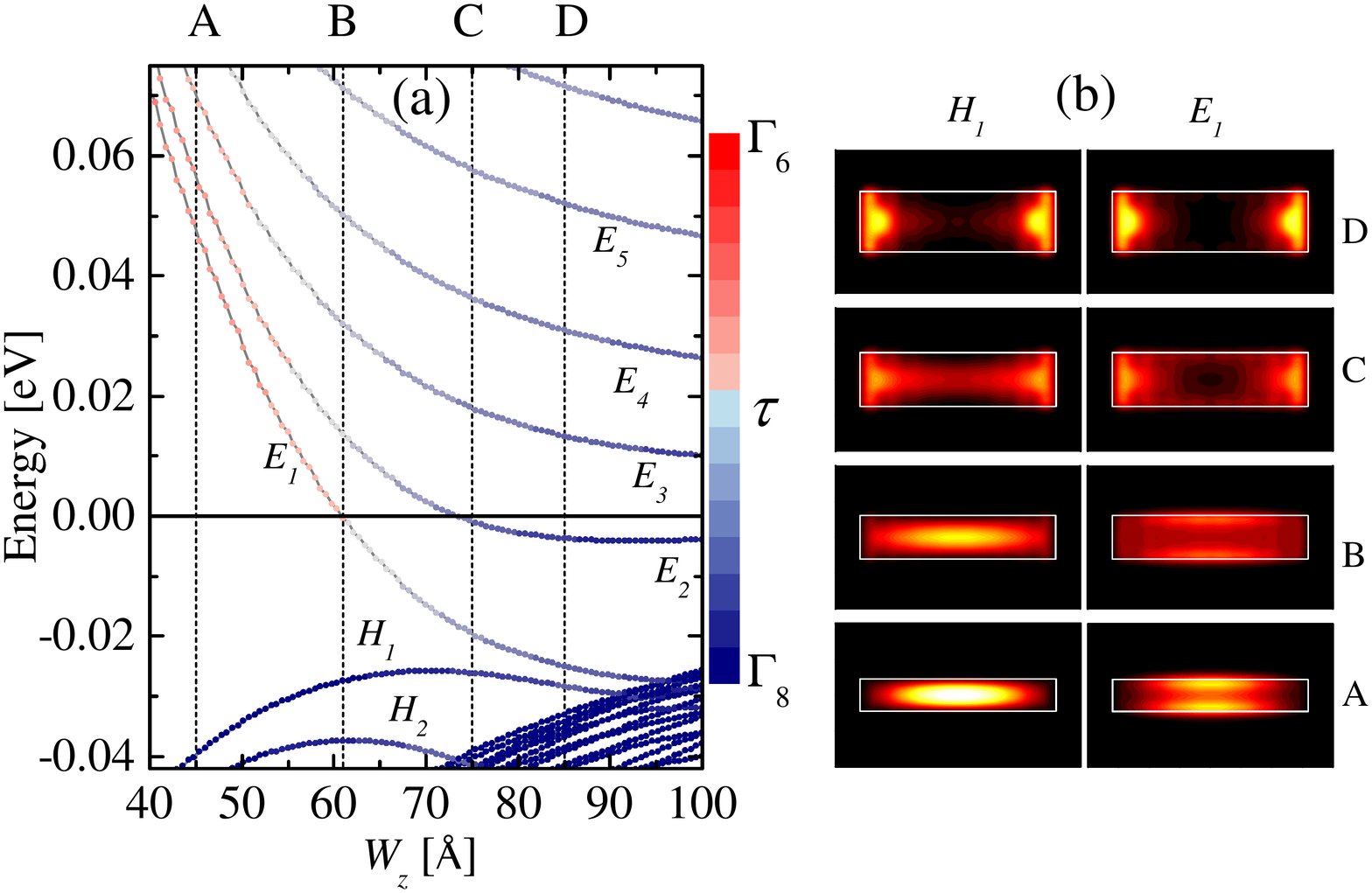}
\caption{(Color online) (a) Subband energies at $k_x=0$ in a thin HgTe/CdTe nanowire with rectangular cross-section (without groove) as a function of its thickness along the $z$ axis, $W_z$. All the energy levels shown in the spectrum have a color scale indicating their main character (from 100$\%$ $\Gamma_6$ to 100$\%$ $\Gamma_8$). (b) The probability density, $|\psi(\mathbf{r})|^2$, for the states $H_1$ and $E_1$ at the same selected $W_z$ values indicated in (a): A ($4.5$ nm), B ($6.1$ nm), C ($7.5$ nm) and D ($8.5$ nm).}\label{fig-2}
\end{figure}

\begin{figure}[t]
\includegraphics[trim=0cm 0cm 0cm 0cm, clip=true,width=0.5\textwidth]{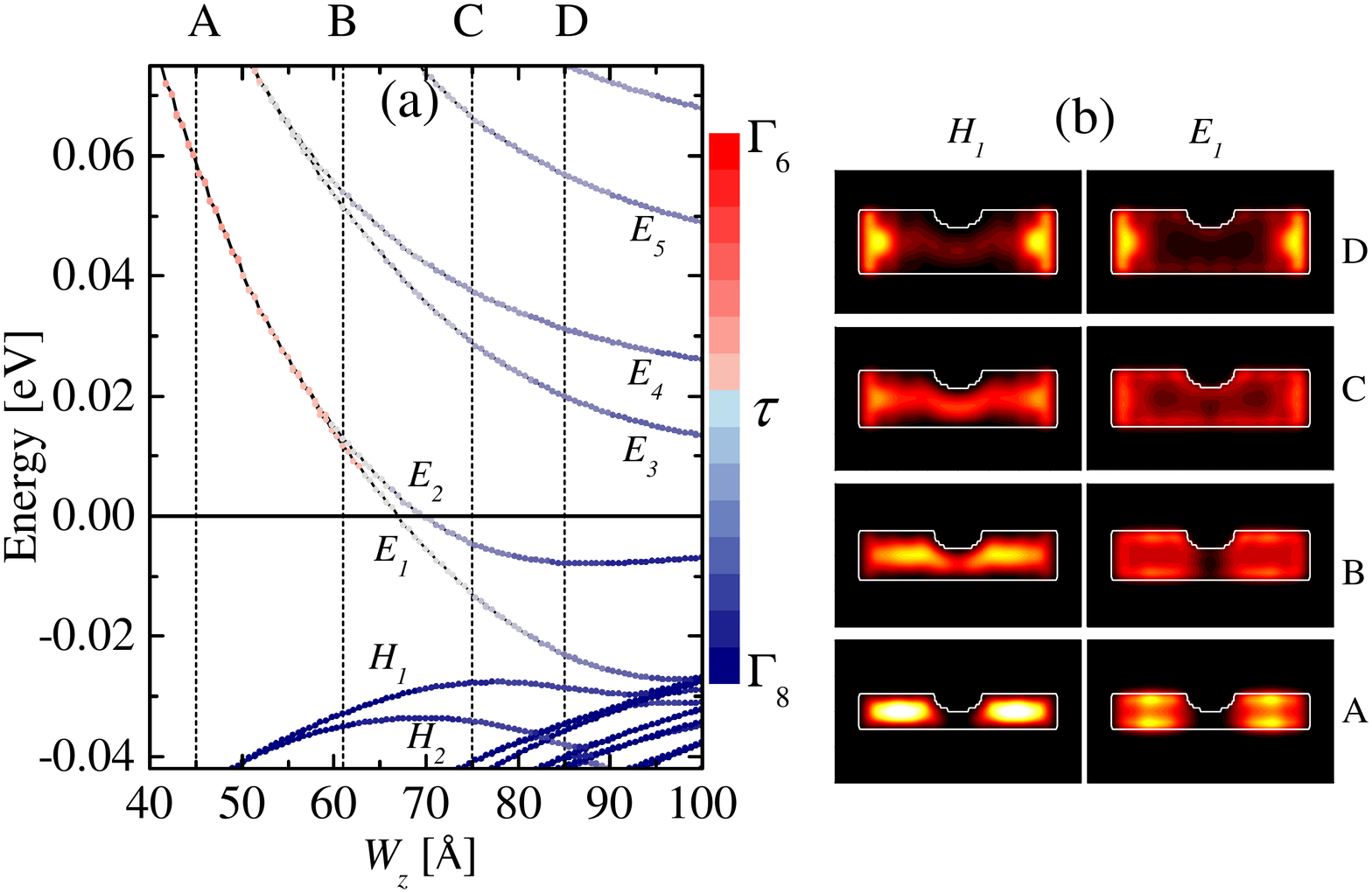}
\caption{(Color online) The same as in Fig.~\ref{fig-2} but for the case of the nanowire with the longitudinal groove as part of its cross-section.}\label{fig-3}
\end{figure}

As discussed in the introduction, quantum well systems based on the combination HgTe/CdTe --and their alloys-- are extremely sensitive to the well thickness. Therefore, it is convenient to begin by analyzing the structure of the subbands in the nanowire as a function of its thickness $W_z$  --in a nanowire with the cross-section shown in Fig.~\ref{fig-1} and also {\it without} the longitudinal groove-- in order to show some basic features before studying in detail the effect of the external electric field and the Rashba splitting in our system. This first step will allow us to establish some definitions that will be useful in the rest of the paper.

The figure~\ref{fig-2}(a) shows the dependence of the energy spectrum versus $W_z$ of the rectangular nanowire, calculated at $k_x=0$. Since the spectrum is calculated at the $\Gamma$ point and there is no external magnetic field present, each of the states shown in the figure is two-fold spin degenerate (Kramers doublet), so the label $s$ is absent in the labels used. The color scale applied to the spectrum shows the main composition of these states in terms of the symmetry groups $\Gamma_6$ and $\Gamma_8$ (for the energy range considered here the contribution of $\Gamma_7$ is negligible and was excluded from the analysis). The parameter $\tau$ (spanning the range $[-1:1]$) associated with each color is given by:
\begin{eqnarray}\label{eq:spinconductance}
\tau &=& \sum_{\alpha=1}^{2}\int_{\Omega}d^2\bm{r}|\chi_{\alpha}(\bm{r})|^2 - \sum_{\alpha=3}^{6}\int_{\Omega}d^2\bm{r}|\chi_{\alpha}(\bm{r})|^2\text{.}\nonumber\\
\end{eqnarray}

At first sight, the qualitative behavior of the energy levels $E_1$ and $H_1$ is slightly similar to that of a quantum well ~\cite{Winkler2012, Sengupta2013}, especially for smaller thicknesses $W_z$. In the latter, the BIA opens a small gap --of approximately $2.9$ meV-- where the crossing between levels $E_1$ and $H_1$ should occur. In our case, however, the extra confinement along the $y$ axis modifies the dependence on $W_z$ in two ways: in first place on the energy levels, shifting the crossing between $E_1$ and $H_1$ to values of $W_z$ over $9$ nm. The position of the crossing in the $W_z$ axis is strongly related to the width $W_y$ of the nanowire (see the Fig. 1 of the supplementary material document). For the width considered here, this crossing coincides with a dense group of $\Gamma_8$ valence states that shifts upwards. In second place, on the composition of the states. Contrary to what is observed in the quantum well, where the crossing also coincides with the $\Gamma_6\rightarrow\Gamma_8$ inversion, here $E_1$ goes from $\Gamma_6$ to $\Gamma_8$ long before the crossing with $H_1$ (around $6.1$ nm in this case). This transition is not abrupt but rather occurs progressively. In the case of states $E_n > E_1$, the $\Gamma_6\rightarrow\Gamma_8$ transition occurs even at a smaller thicknesses.

The fact that the bulk effective masses inside and outside the HgTe/CdTe quantum well are of opposite sign makes some eigenstates tend to locate at the interfaces of the well. In the case of the nanowire, it is possible to observe a similar effect. Note in Fig.~\ref{fig-2}(b) that, as the thickness $W_z$ is increased, the progressive hybridization of $E_{1}$ with $H_{1}$ gives place to the formation of states mainly located at the interfaces perpendicular to the $y$-axis. For the narrowest wire, the shape of the wave function is different in $H_1$ and $E_1$. Nevertheless, as $W_z$ is increased, both states become more and more similar in shape, in line with the progressive reduction of the energy gap between these. 

The presence of a longitudinal groove modifies significantly the energy spectrum and its dependence on the nanowire thickness. As one might expect, in Fig.~\ref{fig-3}(a) we observe that the differences --both quantitative and qualitative-- between this spectrum and those shown in Fig.~\ref{fig-2}(a) become more and more remarkable as $W_z$ decreases. The most obvious difference is the progressive approach between the $E_1$ and $E_2$ doublets (also between $E_3$ and $E_4$ and so on) as $W_z$ reach $4$ nm, where they are almost degenerate. This behavior can also be observed between $H_1$ and $H_2$. 

The origin of the strong approach between the first two Kramers doublets becomes clear by setting our problem in terms of a double potential well, i.e.,  two quantum wells separated by a potential barrier right in the middle: if the barrier between the two wells tends to infinity, then the two lowest states are degenerate. This zero energy difference can be broken when the barrier between the wells is thin enough to allow tunneling. In our case, each one of the lowest (highest) conduction (valence) Kramers doublets $E_1$ and $E_2$ ($H_1$ and $H_2$) are split in two parts spatially separated due to the longitudinal groove [see Fig.~\ref{fig-3}(b)]. However, this geometrical configuration still permits a small interaction between the two regions located left and right of the groove, which is the origin of the small energy difference between these states. Obviously, this ``tunneling splitting'' increases as we increase $W_z$. 

\begin{figure}[t]
\begin{centering}
\includegraphics[trim=0cm 0cm 0cm 0cm, clip=true,width=0.4\textwidth]{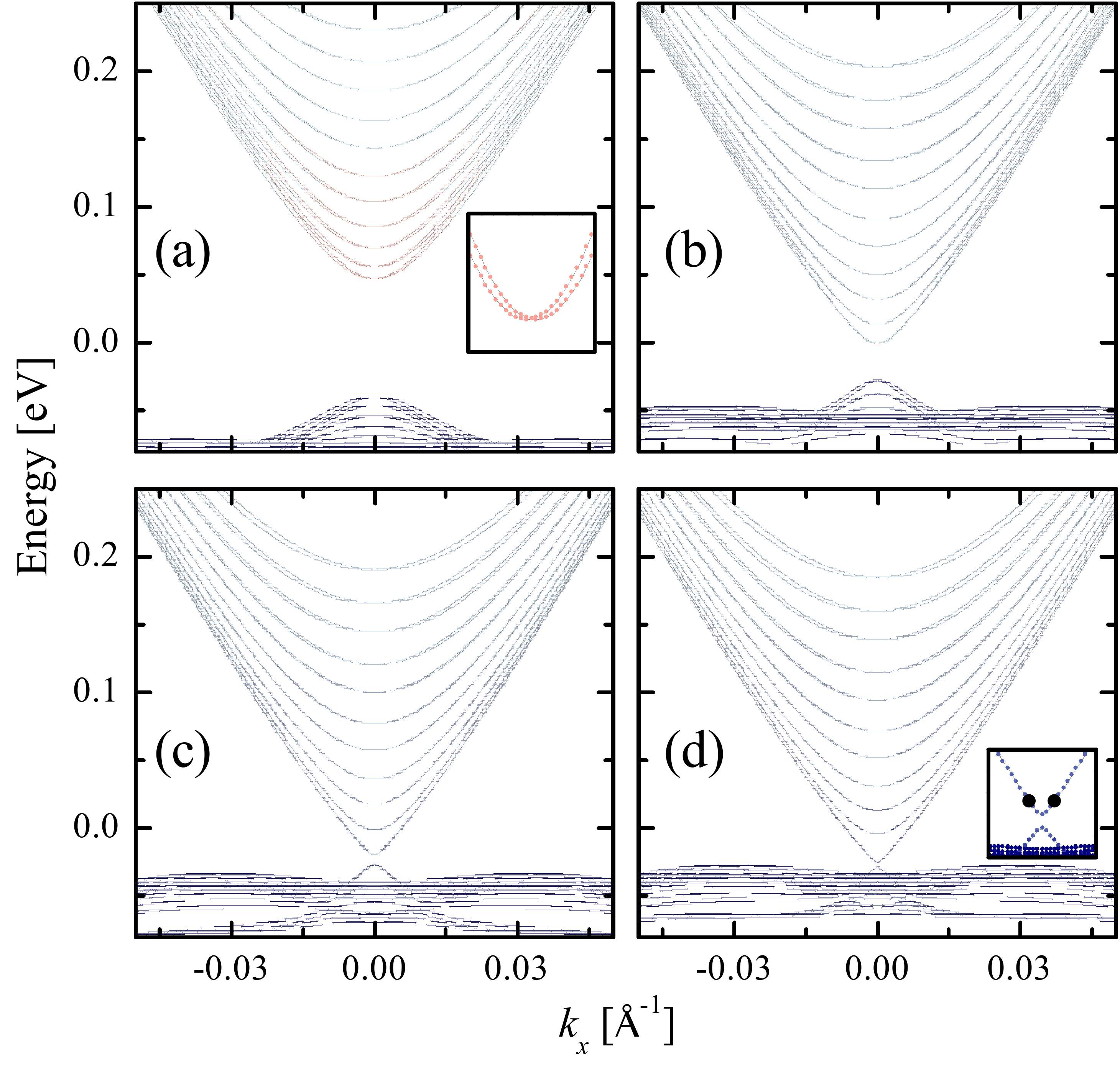}\\
\includegraphics[trim=0cm 0cm 0cm 0cm, clip=true,width=0.35\textwidth]{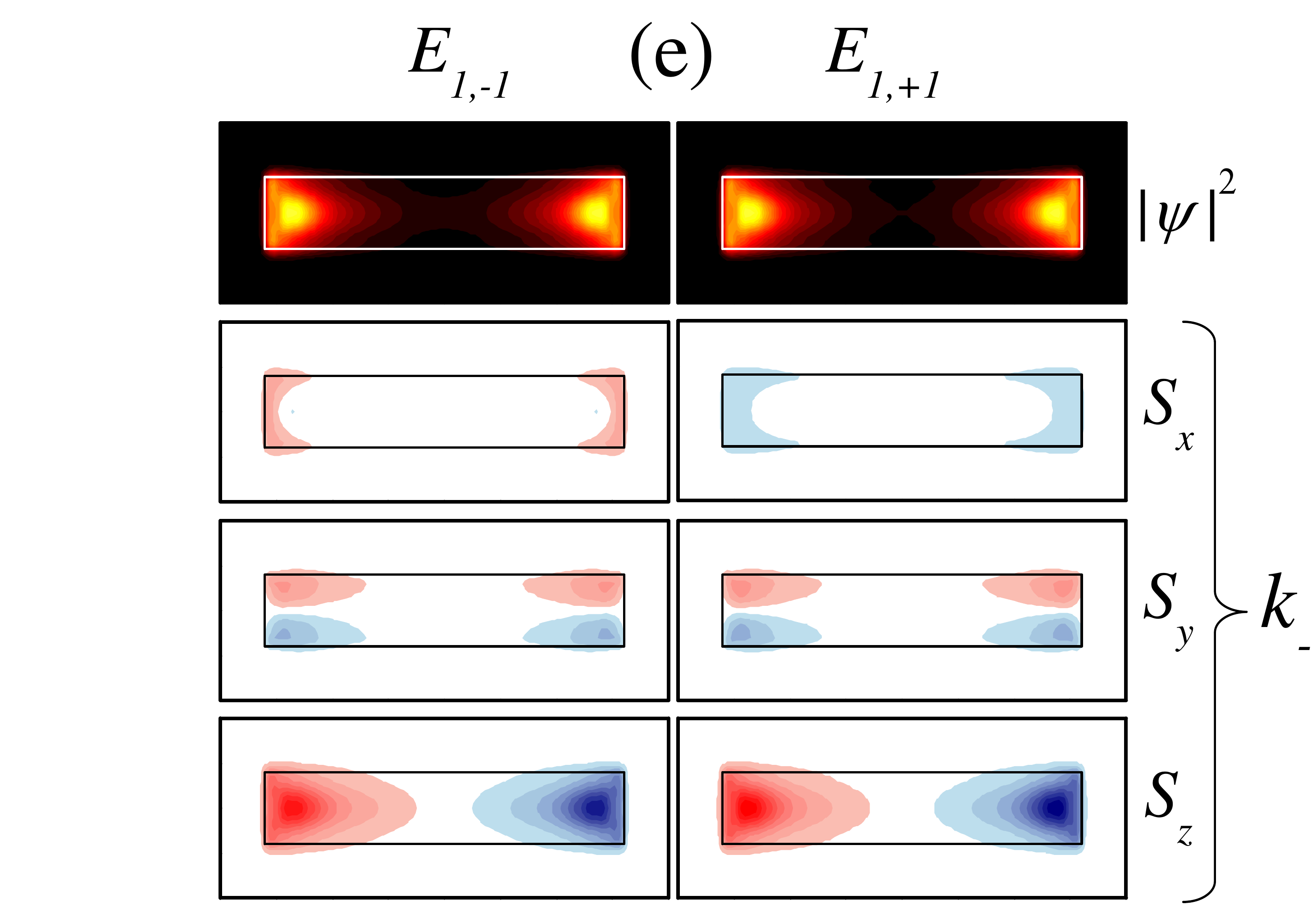}\\
\includegraphics[trim=0cm 0cm 0cm 0cm, clip=true,width=0.35\textwidth]{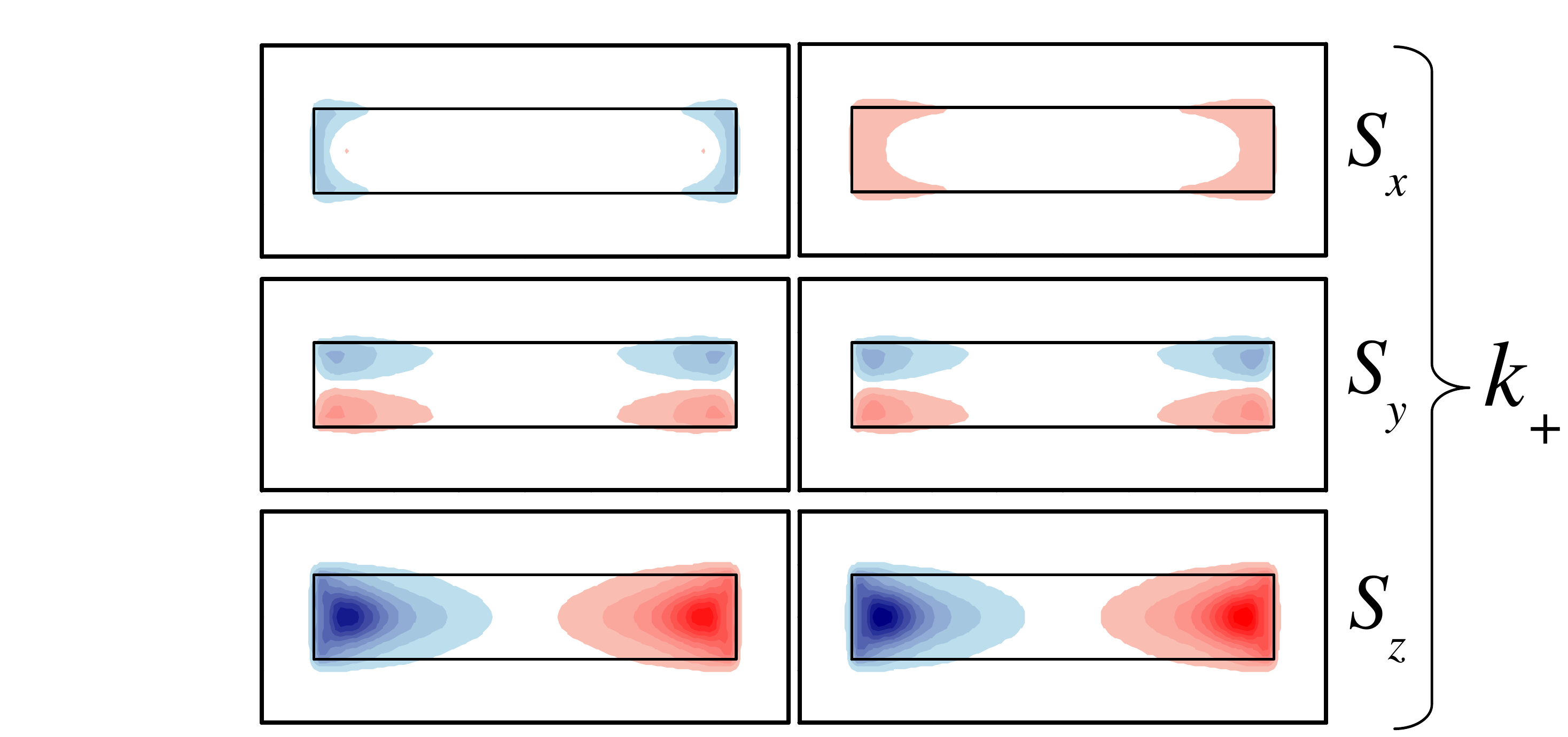}\\
\end{centering}
\caption{(Color online) (a-d) Calculated subband energy dispersions of the nanowire with the rectangular cross-section for different thicknesses: (a) $W_z=4.5$ nm, (b) $W_z=6.1$ nm, (c) $W_z=7.5$ nm and (d) $W_z=8.5$ nm. The color scale has the same meaning that in Figs.~\ref{fig-2} and ~\ref{fig-3}. (e) Probability densities and cartesian components of the spin density (both in arbitrary units) for the states $E_{1,\pm1}$, at $k_x=-0.001$ \AA$^{-1}=k_-$ and $k_x=+0.001$ \AA$^{-1}=k_+$, of the $8.5$ nm thick nanowire [indicated by two thick black dots on the inset in (d)]. Regarding the spin densities, the dark blue color corresponds to negative values while the red correspond to positive values. Here, only the doublet $E_1$ has been shown since $H_1$ present identical qualitative features.}\label{fig-4}
\end{figure}

\begin{figure}[t]
\includegraphics[trim=0cm 0cm 0cm 0cm, clip=true,width=0.4\textwidth]{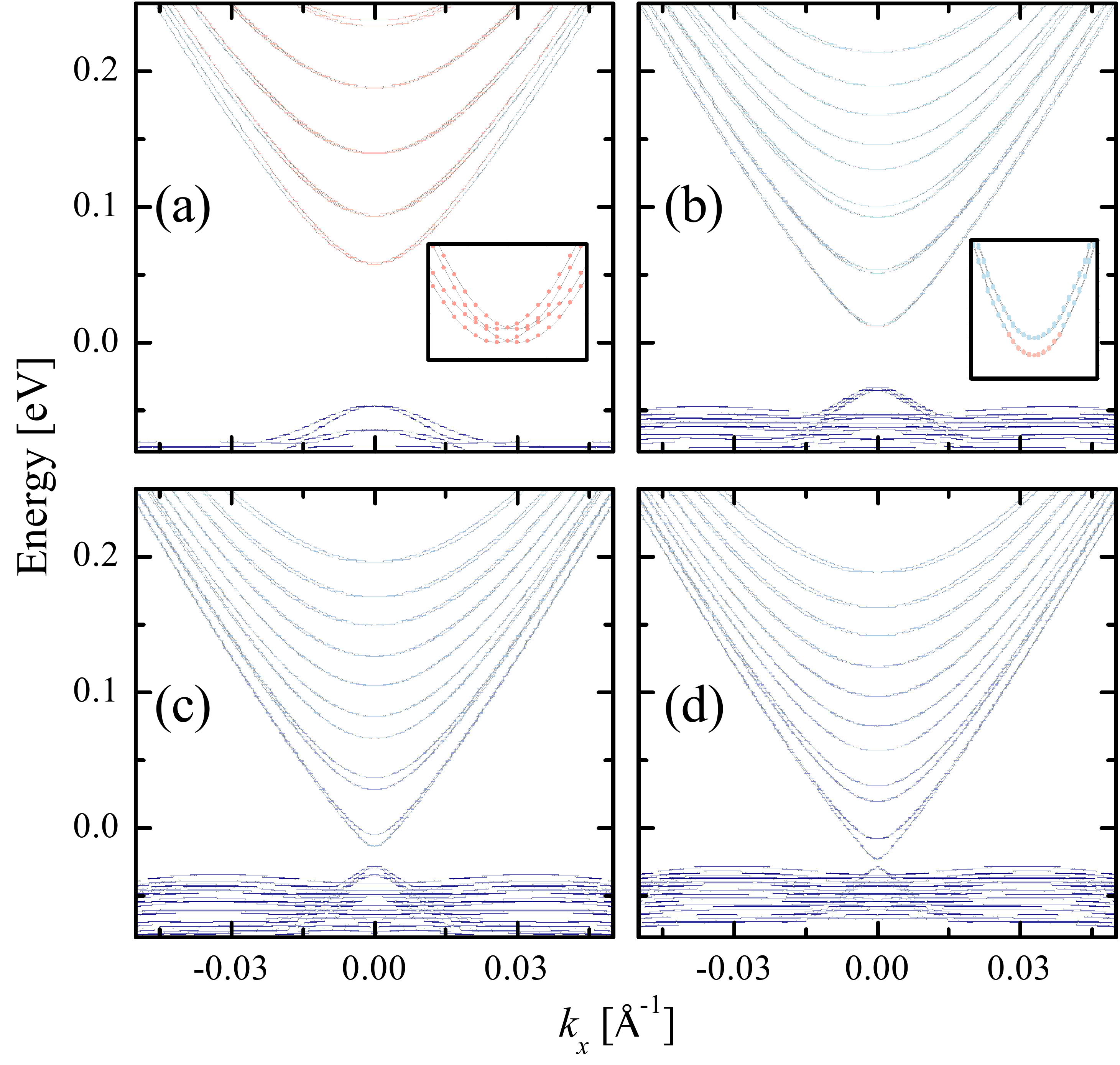}\\
\caption{(Color online) The same as Figs.~\ref{fig-4}(a-d) but for the case of the nanowire with the longitudinal groove.}\label{fig-5}
\end{figure}

In Fig.~\ref{fig-4}, the subband structures of the rectangular quantum wire, for the four thicknesses $W_z$ considered previously, are shown. As expected for the HgTe, note that the dispersion of the conduction subbands is far from being parabolic. 
In the case of the rectangular nanowire, as $W_z$ is increased, the dispersion tends to resemble more and more a Dirac cone while the gap between the conduction and valence subbands decreases, as we could see in Fig.~\ref{fig-2}(a). Also, as $W_z$ is increased, the band inversion occurs mainly around the $\Gamma$ point and for the few lowest conduction subbands.
For values of $k_x$ far enough from zero (not shown here) the wave functions associated with these subbands change from $p$-type to $s$-type again as these recover the orbital character associated to a normal order of the Energy bands~\cite{Sengupta2013, Teo2008}. In all cases, the BIA spin splitting in the conduction subbands is so small that it is impossible to observe it at the scale of the figures (on the inset in Fig.~\ref{fig-4}(a), a close-up of the subbands $E_{1,\pm1}$ just around $k_x=0$ is shown). However, due to the confinement along the $y$ axis, the BIA is manifested not only in the splitting of the subbands as a function of the momentum, but in the modification of the subbands energy at $k_x=0$. Except for the splitting, the effect does not seem to be important in the case of the conduction subbands but in the case of the valence bands it is (not shown here).

Finally, by direct visual inspection of the wave functions and spin densities one can check that the states in the subbands $E_{1,\pm1}$ and $H_{1,\pm1}$, that form the Dirac-type cones just around $k_x=0$ in Fig.~\ref{fig-4}(d) (see the inset), behave as edge states. These are located very close to the interfaces (in this case, those perpendicular to the $y$ axis). On the other hand, the only  non-zero average spin projection is the component $\langle S_x \rangle$, i.e. parallel to the axis of the wire and to the interfaces where these states are located. The latter is evident by observing, for example, the probability density $|\psi(\bm{r})|^2$ and the components of spin density $S_i(\bm{r})=\psi^*(\bm{r})\hat{S}_i\psi(\bm{r})$ ($i=x,y,z$) in Fig.~\ref{fig-4}(e) for the states $E_{1,\pm1}$ at $k_x=\pm0.001$ \AA$^{-1}$. In both subbands, the $S_y$ and $S_z$ components have a symmetric texture that translates into a zero average for each one of these. As for its spatial distribution, the local behavior of $S_z(\bm{r})$ allows to verify the spin-momentum correlation characteristic of the helical edge states~\cite{Maciejko2011} in the region coincident with the maxima of the $|\psi(\bm{r})|^2$. In each branch of the doublet $E_1$, each edge holds two counter-propagating modes (with momentum $k_-$ and $k_+$) whose spin components $S_z$ are antiparallel. In principle, since time reversal symmetry (TRS) is preserved, these states are topologically protected against backscattering from time-reversal invariant potentials (e.g., non-magnetic impurities). However, this is rigorously true for a single edge or for two edges with no interaction between them. For the nanowire width considered here ($W_y = 80$ nm), the overlap between the tails of the spatially separated edge states, although small, is not negligible. In this situation, in presence of a scattering source, the probability that an electron moving along one of the edges performs a backscattering by jumping to the opposite edge would not be zero. This aspect should  be ruled out by increasing the width of the nanowire enough to make this overlap negligible. 

Moving away from $k_x=0$, the progressive approach of the edge states to the bulk-type subbands results in a decrease in the accumulation of the probability density at the edges (see Figs. 2(a) and 2(b) of the supplementary material document).  Far from $k_x=0$, the result agrees with what would be expected in a BIA spin-orbit coupled nanowire with normal band ordering.    

In the case of a nanowire with the longitudinal groove (see Fig.~\ref{fig-5}), we observe that for both $W_z=4.5$ nm and $W_z=6.1$ nm, the separation between the two lowest (highest) Kramers doublets of the conduction (valence) subbands is so small that it is difficult to distinguish them from the scale we have used in the figures. In the insets of Fig.~\ref{fig-5}(a) and Fig.~\ref{fig-5}(b) a close-up of the lowest energy conduction subbands $E_{1,\pm1}$ and $E_{2,\pm1}$ is shown around the $\Gamma$ point: note in the inset in (a) that both $E_{1,\pm1}$ and $E_{2,\pm1}$ are mostly $\Gamma_6$ whereas in the inset in (b) only $E_{1,\pm1}$ are $\Gamma_6$. As in Fig.~\ref{fig-4}, just around $k_x=0$ the dispersion of the lowest subbands is approximately parabolic. Then, as $W_z$ increases, the gap between the conduction and valence subbands not only decreases, but also a progressive modification towards a linear dispersion around the $\Gamma$ point is observed. Finally, note in the Fig.~\ref{fig-5} how, as $W_z$ increases, this tunneling splitting occurs exclusively around $\Gamma$. 

\begin{figure}[t]
\includegraphics[trim=0cm 0cm 0cm 0cm, clip=true,width=0.5\textwidth]{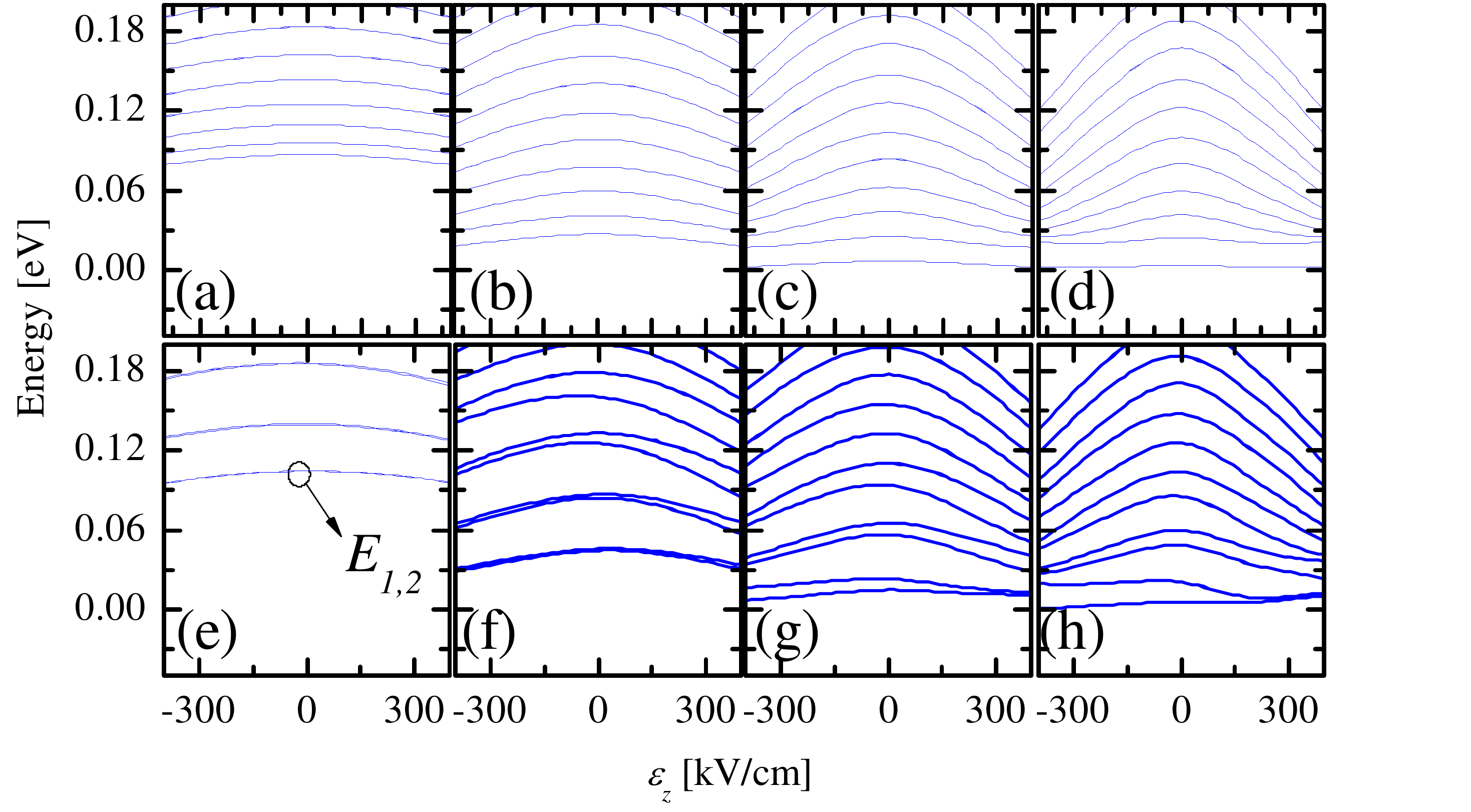}\\
\caption{(Color online) (a-d) Subband energies calculated at $k_x=0$ of the nanowire with the rectangular cross-section as a function of the external electric field strength along the $z$ axis, $\varepsilon_z$, for diferent thicknesses [(a) $W_z=4.5$ nm, (b) $W_z=6.1$ nm, (c) $W_z=7.5$ nm, (d) $W_z=8.5$ nm]. Only the conduction type subbands are shown here. The energy of the lowest conduction subband was
substracted on each level at each value of $\varepsilon_z$, i.e., $E_n-E_1$. (e-h) The same as plots (a-d) but for the case of the nanowire with the longitudinal groove.}\label{fig-6}
\end{figure}

Next, we will study the effects of an external electric field on the energy spectrum at $k_x=0$ first, where Rashba SOC effects aren't present. The figure~\ref{fig-6} shows the dependence of the energy levels of the nanowire --the two shapes considered in this work-- as a function of the electric field strength. Here, only the conduction subbands are shown. Since we are mostly interested in studying the dependence of the relative separation in energy between the subbands, all the energies have been shifted with respect to the highest valence energy level ($H_1$) for each value of $\varepsilon_z$. Thus, the lowest conduction energy level shown in Fig.~\ref{fig-6} set, in fact, the value of the gap between the conduction and valence bands. Apart from the obvious symmetry of the spectrum with respect to the sign of the electric field, the first feature that one can extract from Figs.~\ref{fig-6}(a-d)  is that the separation between $E_1$ and $E_2$ varies very little as the strength of the field increases, whereas the same does not occur in the states with higher energies. In the latter case, a rapid decrease in the energy separation between these states and $E_1$ is observed in nanowires with $W_z\ge6.1$ nm. For $W_z=4.5$ nm, the strong confinement in the same direction as the electric field avoids the displacement of these states associated to the tilt of the confinement profile, making this separation negligible. On the other hand, in the nanowire with the longitudinal groove [Figs.~\ref{fig-6}(e-h)], the behavior of the energy levels as a function of the electric field is different depending on its sign. This difference is important in nanowires with larger thicknesses, whereas in the case of the $4.5$ nm thick nanowire is observed to a lesser extent. In the latter case, the doublets $E_1$ and $E_2$ are separated by the small tunneling splitting previously defined. As it was studied above, the increase of $W_z$ results in an increasing tunneling splitting at $\varepsilon_z=0$. However, in (g) and (h), the tunneling splitting between the first two subbands decreases rapidly by increasing the electric field in the direction parallel to the $z$ axis ($\varepsilon_z>0$) (on the contrary, in the higher energy subbands this effect is not evident). When applying the electric field in this direction, part of the states $E_1$ and $E_2$ move towards the upper interface of the nanowire, just where the groove is located. The combination of the tilt of the confinement profile along $z$ and the presence of the longitudinal groove forces these states to separate spatially, showing again characteristic features of a double potential well and thus reducing the tunneling splitting between them. In addition, one can observe in (g) and (h) that, for $\varepsilon_z>0$, the levels $E_1$ and $E_2$ shift both toward higher energies due to the crossing of the conduction and valence bands. 

\begin{figure}[h]
\includegraphics[trim=0cm 0cm 0cm 0cm, clip=true,width=0.47\textwidth]{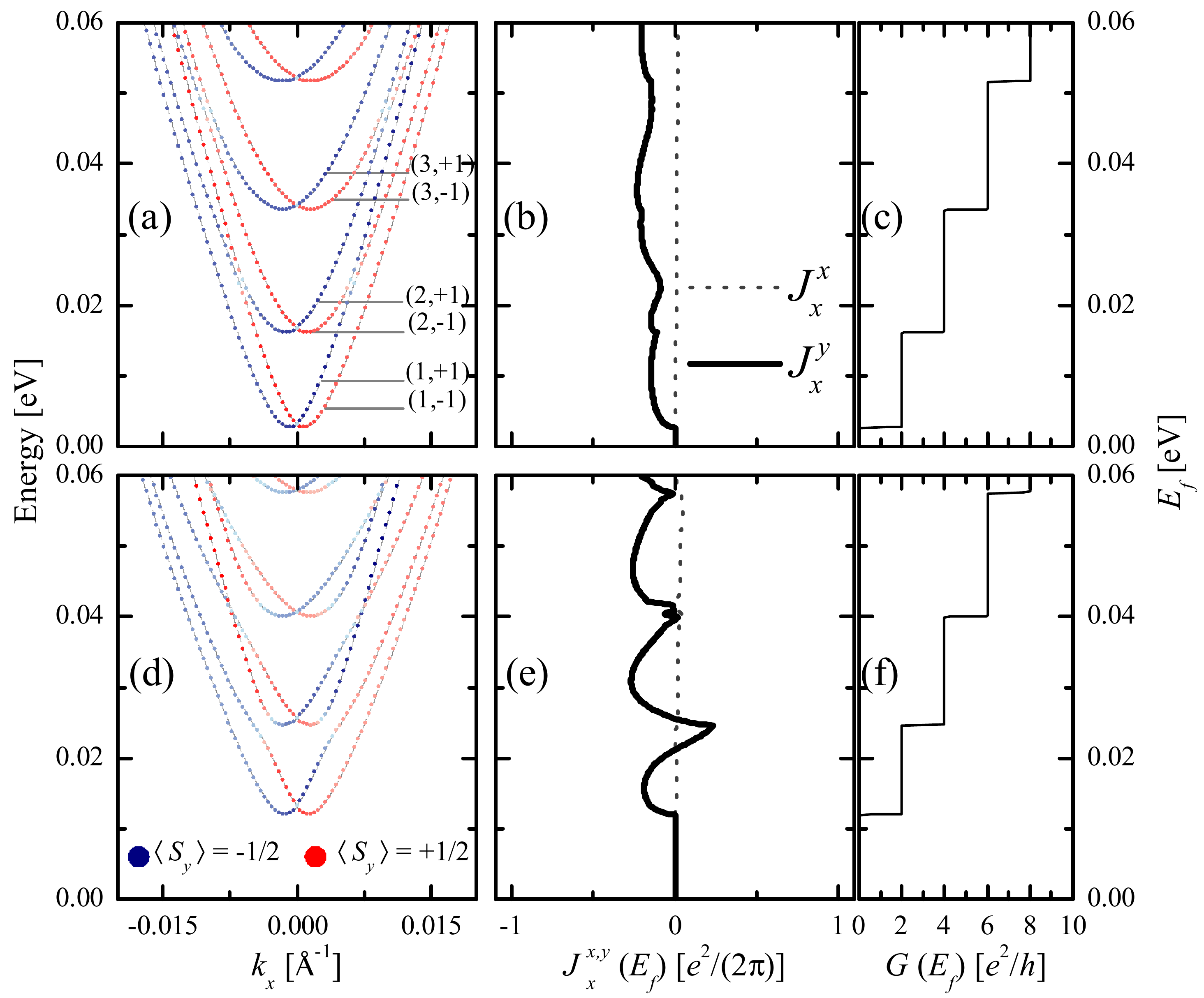}\\
\caption{(Color online) (a) Subband energy dispersions of the $6.1$ nm thick rectangular nanowire at $\varepsilon_z=150$kV$/$cm. The color scale refers to the $y$ component of $k_x$-dependent expectation value of the spin, $\langle S_y \rangle$, which in our case is the dominant component. The labels $(n,s)$ refers to the index of the subband $E_{n,s}$. (b) The $x$ and $y$ components of the spin-polarized conductance as a function of the Fermi level (here, we considered only left-to-right propagation, i.e., only those chanels with positive group velocities). (c) The charge conductance calculated under the same assumptions that in (b). Figs. (d), (e) and (f) describe the same as Figs. (a), (b) and (c), respectively, but at $\varepsilon_z=300$kV$/$cm. }\label{fig-7}
\end{figure}

We now turn to study the effects of the SOC changing our attention to states away from $k_x=0$, where one can expect the occurrence of the Rasha spin splitting. The presence of this splitting has important effects on the energy spectrum as it might lead to the appearance of multiple $k_x$-dependent anticrossings. These anticrossings represent ``hot spots'' in which transitions between opposite-spin states are more probable. In the following, we will study the structure of subbands in the two types of nanowire in the presence of the Rashba splitting for some of the examples shown above, together with the spin and charge conductances associated with these dispersions.  

\begin{figure}[t]
\includegraphics[trim=0cm 0cm 0cm 0cm, clip=true,width=0.47\textwidth]{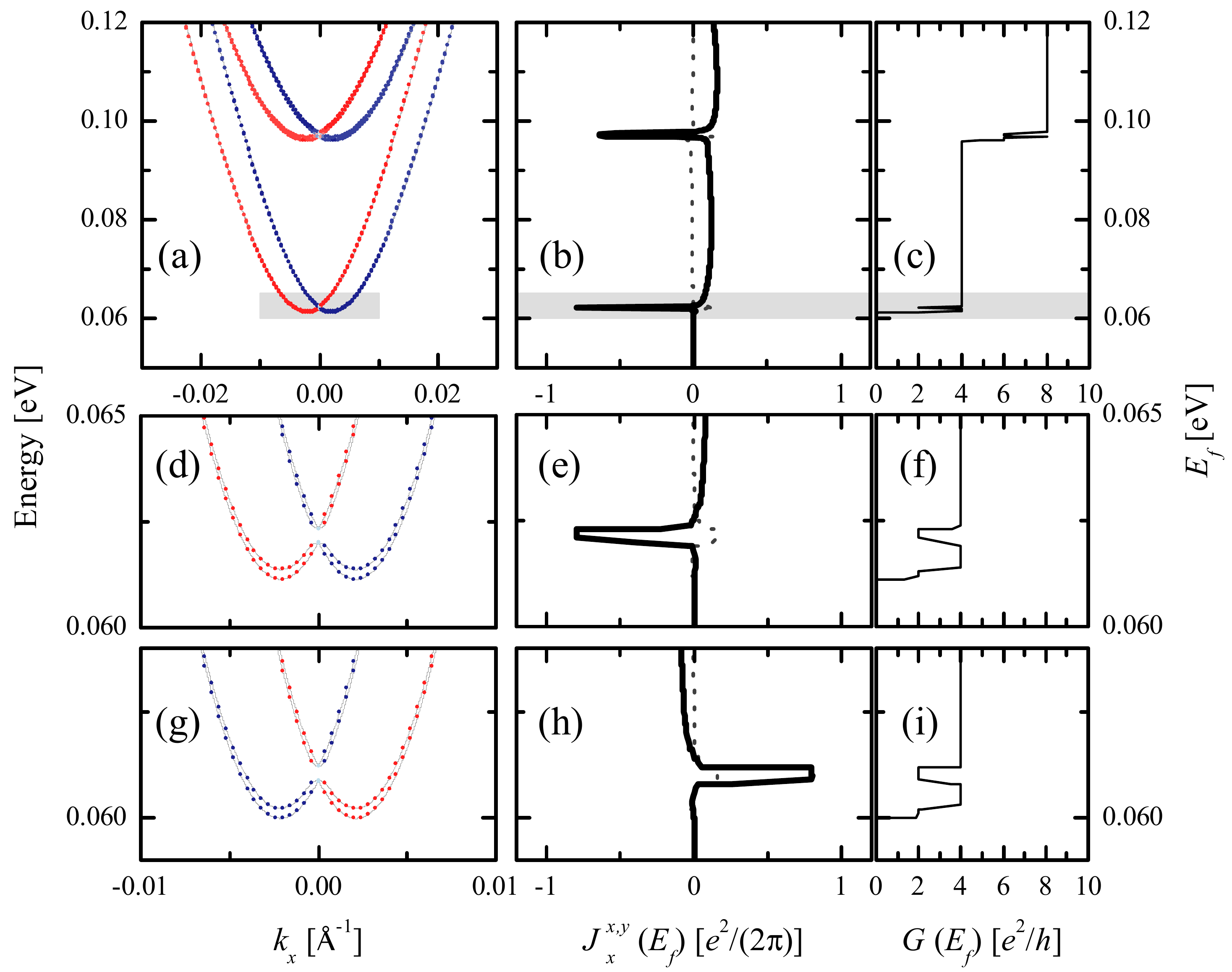}\\
\caption{(Color online) (a) Subband energy dispersions of the $4.5$ nm thick groove-nanowire at $\varepsilon_z=-300$kV$/$cm. (b) The $x$ and $y$ components of the spin-polarized conductance as a function of the Fermi level. (c) The charge conductance as a function of the Fermi level. The color and line-type details are the same that in Fig.~\ref{fig-7}. (d-f) Detail of the shaded areas in figures (a), (b) and (c), respectively. (g-i) The same as figures (d), (e) and (f) but at $\varepsilon_z=+300$kV$/$cm.}\label{fig-8}
\end{figure}

\begin{figure}[t]
\includegraphics[trim=0cm 0cm 0cm 0cm, clip=true,width=0.47\textwidth]{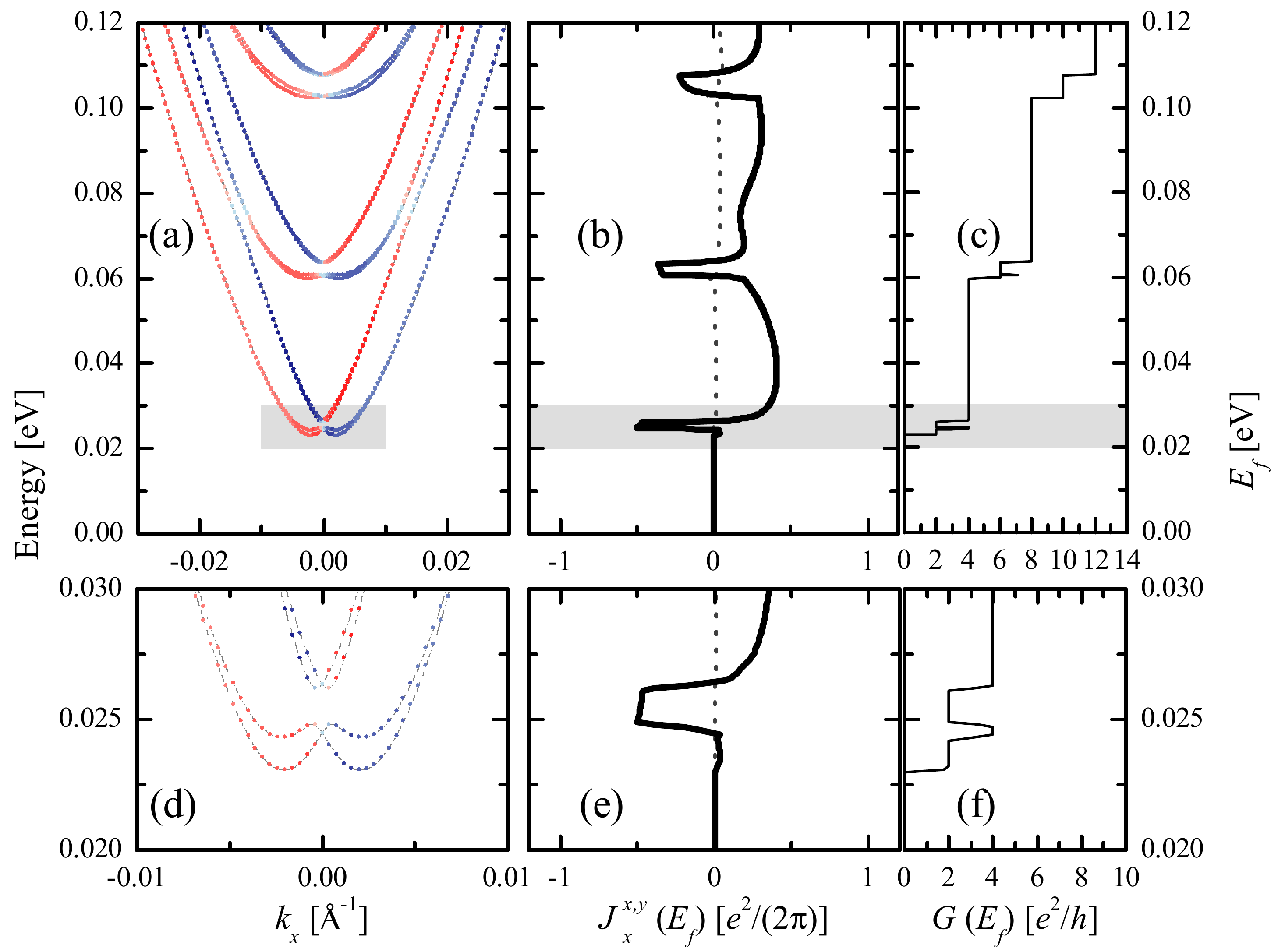}\\
\caption{(Color online) The same as in Fig.~\ref{fig-8} but for a  $6.1$ nm thick groove-nanowire at $\varepsilon_z=-300$kV$/$cm. }\label{fig-9}
\end{figure}

\begin{figure}[h]
\includegraphics[trim=0cm 0cm 0cm 0cm, clip=true,width=0.47\textwidth]{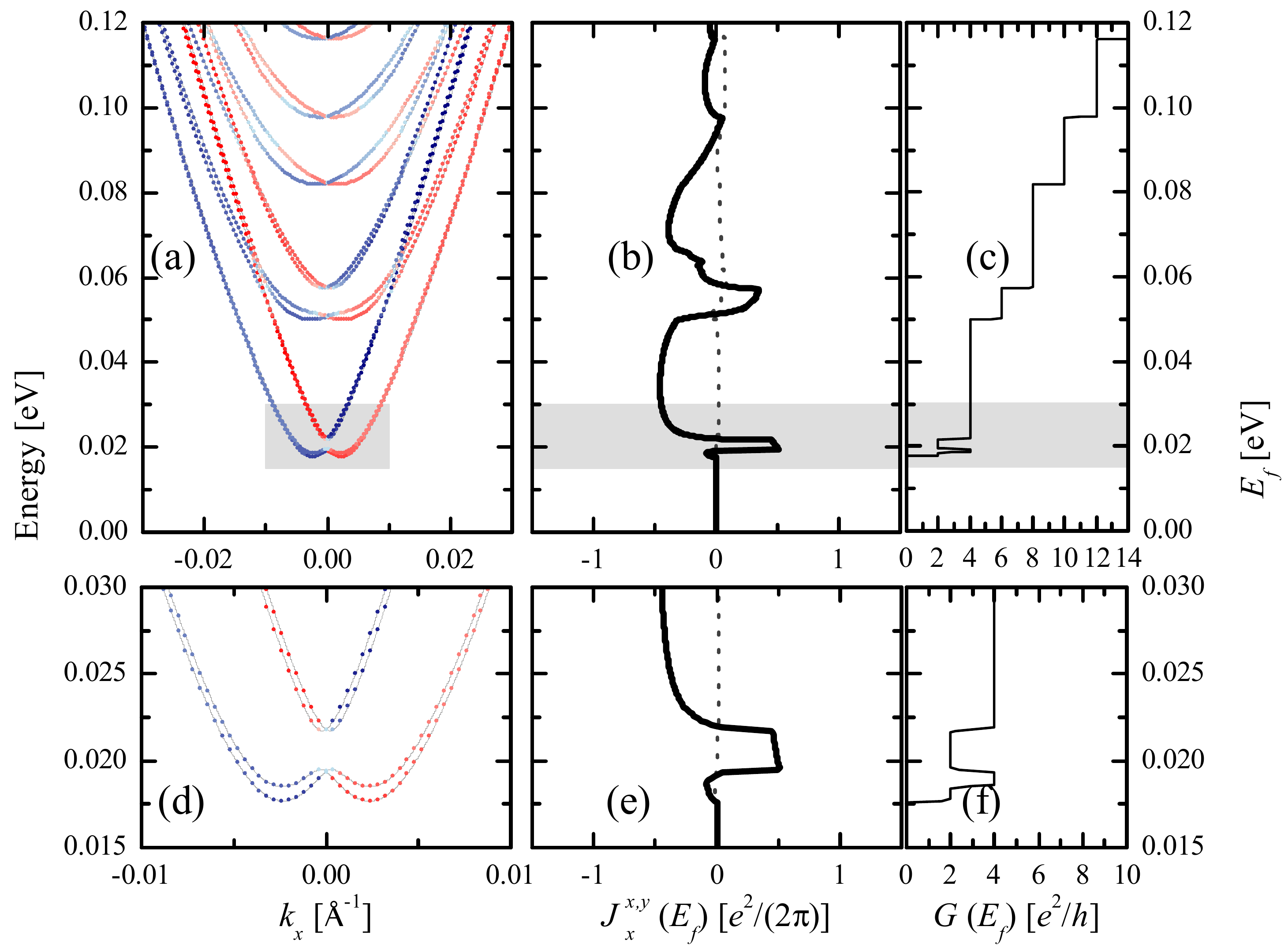}\\
\caption{(Color online) The same as in Fig.~\ref{fig-10} but at $\varepsilon_z=300$kV$/$cm. }\label{fig-10}
\end{figure}

In terms of applications in realistic devices, the lowest energy subbands are those that usually determine the transport properties of the device, so in the following we have focused on these. The Figs.~\ref{fig-7}(a) and~\ref{fig-7}(b) show the dispersion of the lowest conduction subbands of the $6.1$ nm thick rectangular nanowire for two electric field strengths: $\varepsilon_z=150$ kV/cm and $\varepsilon_z=300$ kV/cm, respectively. In addition, the spin and charge conductances are also shown. In the latter, a staircase with step heights in units of $2\times e^2/h$ is obtained (typical of quantum wires, being $e^2/h$ the quantum for charge conductance). The first feature to note in Fig.~\ref{fig-7}(a) is that the spin projection $\langle S_y \rangle$ of the subbands changes very little except around the anticrossings, where hybridization between adjacent subbands with opposite spin projections results in a strong reduction of the magnitude of these. We have chosen to represent this projection in the figure --perpendicular to both the electric field and the nanowire axis-- because it turns out to be the dominant. The $\langle S_z \rangle$ component is zero across the range of energies studied while the component $\langle S_x\rangle$ --parallel to the wire axis-- although not zero, is very small in comparison with $\langle S_y\rangle$ The latter can be inferred directly from Fig.~\ref{fig-7}(b). This result coincides with what might be expected from the Rashba effect~\cite{Ganichev2004}. 

The calculated spin conductance for this band structure is mainly polarized along $y$ and is negative. The sign of this conductance comes from the combination of the sign of the spin projection and the sign of the group velocity in Eq.~\eqref{eq:spinconductance}. Also, the sign of the conductance will be linked in this case to the sign of the electric field (for this symmetric wire a negative $\varepsilon_z$ gives an identical spin conductance but with an opposite sign). On the other hand, although $J_x^y$ does not change its sign and is approximately regular within the range of energies considered, it is possible to see small variations in its value. The position of these shallow dips  in $J_x^y$ coincides with the energies where the anticrossings of the band structure occur. Since these anticrossings are still far from $k_x=0$, the change in the curvature of the dispersion of the subbands induced by these --and consequently, in the group velocities-- is not important enough to affect appreciably the shape of the spin conductance. This situation change completely as the electric field strength is increased up to $300$ kV/cm [Figs.~\ref{fig-7}(d-f)]: in this case, the Rashba splitting turns out to be large enough to cause the anticrossings to move towards the $\Gamma$ point. The hybridization between the subbands $E_{1,+1}$ and $E_{2,-1}$ makes $E_{1,-1}$ the only subband that contributes to $J_x^y$ (with $\langle S_y \rangle > 0$). Therefore, a change of sign in $J_x^y$ --from negative to positive values-- occurs. Since the anticrossing is just below the $E_{2,\pm1}$ subbands, as just as $E_f$ is increased two new propagation modes appear. This time, the value of the component $\langle S_y \rangle$ of the states with positive group velocities that participate in the propagation is quite close to $-1/2$, so they contribute to $J_x^y$ in such a way that the sign of the latter change to negative values again.    

We have verified that the presence of a longitudinal groove in a flat nanowire gives rise to a kind of tunneling splitting between  some of the subbands (following the analogy of the double potential well). Depending of the deepness of the groove and the thickness of the nanowire, that tunneling splitting may be comparable or smaller to the Rashba spin splitting for moderate electric fields, which may produce interesting effects in the subbands structure and,  consequently, in the spin and charge transport properties. Analogously to that shown in Fig.~\ref{fig-7}, the figure~\ref{fig-8} shows the dispersion of the first lowest conduction subbands in a $4.5$ nm thick groove-nanowire at $\varepsilon_z=-300$ kV/cm (in particular, Figs.~\ref{fig-8}(g-i) correspond to the case $\varepsilon_z=+300$ kV/cm). Unlike the rectangular nanowire, the small energy difference between the doublets $E_{1,\pm1}$ and $E_{2,\pm2}$ (and between $E_{3,\pm1}$ and $E_{4,\pm2}$) at $k_x=0$ and the large Rashba splitting generated by the electric field produce a strong shift of the anticrossings toward the $\Gamma$ point. The resulting modification in the curvature of the subbands involved in the anticrossings opens a ``mini gap'' that is qualitatively similar to the zero-momentum splitting induced by an external magnetic field (Zeeman splitting); with the particularity that in our case, the splitting does not occur between two subbands with opposite spins but between pairs of subbands. In the momentum interval between the $\Gamma$ point and the anticrossing, the subbands $E_{n,\pm1}$ have spin projections with opposite signs. This change after the anticrossing so, within the energy interval spanned by the mini gap, the average spin $\langle S_y \rangle$ of both subbands has the same sign. This is important if we consider that the group velocities of both subbands also have the same sign within the energy range spanned by the mini gap. As a result, both subbands contribute constructively to the spin conductance. As it can be seen in Fig.~\ref{fig-8}(b) [\ref{fig-8}(e) and \ref{fig-8}(h)], the spin conductance shows a series of well-defined and narrow pulses coincident with the position of the mini gaps. The polarization of these pulses is close to unity for two propagation modes. This is equivalent to a $y$-component spin projection close to $\pm1/2$ for each mode. Note also that the opening of one of these mini gaps manifests itself as a dip in the charge conductance.

Despite the asymmetry in the shape of the nanowire, the fact that one can observe almost identical results --qualitatively speaking-- for two opposite directions of the electric field should not be surprising considering our previous analysis of the energy spectrum (at $k_x=0$) as a function of the field strength. However, this situation changes as the thickness of the nanowire increases. As it can be seen in Figs.~\ref{fig-9} and~\ref{fig-10}, the increase of the tunneling splitting at $k_x=0$ gives place to an increase in the width of the mini gaps mentioned above. Nevertheless, a higher density of subbands interacting between them due to the strong Rashba splitting and the anticrossings give rise to a progressive loss of polarization and to a more irregular profile of the spin conductance as $E_f$ is increased. 

In conclusion, in the framework of an eight-band envelope function theory, we performed a theoretical study of the band structure and spin-related properties of a HgTe/CdTe nanowire with a rectangular cross-section having a longitudinal groove. The characterization of the band structure has allowed us to verify the presence of edge states in the lowest (highest) conduction (valence) subbands. The emergence of these states coincides with the $\Gamma_6-\Gamma_8$ band inversion when the thickness of the nanowire is increased. Similarly to the case of a $[001]$-oriented quantum well formed by the same material combination, the existence of these edge states is limited to the nearest neighborhood of the $\Gamma$ point. For the two types of nanowire studied in this work, those states are localized at the interfaces perpendicular to the $y$ axis, at least for the range of thicknesses considered here. The helicity of these edge states, characteristic of quantum spin Hall systems, has been verified for the rectangular nanowire. However, since there is a not negligible overlap between the  edge states of opposite edges,  the question about topological protection against backscattering is open in this particular case. A more detailed study about transport properties (including scattering sources) in this system would be required. 

On the other hand, we found that the interplay between the Rashba SOC generated by an external electric field and this particular geometry of the nanowire has a significant impact on the electronic subband structure and allows the modulation and control of a polarized spin conductance without the need of any external magnetic field. This last might be helpful considering that electric fields are more easily accessible and controllable in genuine electronic devices at the nanoscale than magnetic fields. Regular pulses of the spin conductance --as a function of the Fermi level-- with a well-defined polarization and with different widths might be obtained by varying some structural parameters of the nanowire, e.g., for a nanowire with a small thickness and considering a shallower groove it might be possible to increase the width of these spin conductance pulses without losing their strong polarization or regularity. We think that similar effects might be extrapolated to other quantum wires with different shapes and with a different distribution --and number-- of longitudinal grooves. This may allow, in principle,  more room for manipulation regarding the handling of these spin-conductance pulses and their polarization.

\section{Acknowledgements}\nonumber

I would like to thank Eugene Krasovskii for his useful comments and criticisms. This work was partially supported by the Project FIS2016-76617-P of the Spanish Ministry of Economy and Competitiveness MINECO.

\end{document}